\documentclass[preprint,12pt]{elsarticle}

\usepackage{amssymb}
\journal{Applied Mathematical Modelling}

\begin{document}

\begin{frontmatter}

%% Title, authors and addresses

%% use the tnoteref command within \title for footnotes;
%% use the tnotetext command for the associated footnote;
%% use the fnref command within \author or \address for footnotes;
%% use the fntext command for the associated footnote;
%% use the corref command within \author for corresponding author footnotes;
%% use the cortext command for the associated footnote;
%% use the ead command for the email address,
%% and the form \ead[url] for the home page:
%%
%% \title{Title\tnoteref{label1}}
%% \tnotetext[label1]{}
%% \author{Name\corref{cor1}\fnref{label2}}
%% \ead{email address}
%% \ead[url]{home page}
%% \fntext[label2]{}
%% \cortext[cor1]{}
%% \address{Address\fnref{label3}}
%% \fntext[label3]{}

%\title{BED-LOAD PREDICTION IN GAS-LIQUID-SOLID STRATIFIED FLOWS IN HORIZONTAL DUCTS}
\title{PREDICTION OF THE BED-LOAD TRANSPORT BY GAS-LIQUID STRATIFIED FLOWS IN HORIZONTAL DUCTS \tnoteref{label_note_copyright} \tnoteref{label_note_doi}}

\tnotetext[label_note_copyright]{\copyright 2016. This manuscript version is made available under the CC-BY-NC-ND 4.0 license http://creativecommons.org/licenses/by-nc-nd/4.0/}

\tnotetext[label_note_doi]{Accepted Manuscript for Applied Mathematical Modelling, v. 37, p. 5627-5636, 2013, 10.1016/j.apm.2012.11.013}

%% use optional labels to link authors explicitly to addresses:
%% \author[label1,label2]{<author name>}
%% \address[label1]{<address>}
%% \address[label2]{<address>}

\author{Erick de Moraes Franklin}

\address{Faculdade de Engenharia Mec\^anica - Universidade Estadual de Campinas\\
e-mail: franklin@fem.unicamp.br\\
Rua Mendeleyev, 200 - Campinas - SP - CEP: 13083-970\\
Brazil}

\begin{abstract}
%% Text of abstract
Solid particles can be transported as a mobile granular bed, known as bed-load, by pressure-driven flows. A common case in industry is the presence of bed-load in stratified gas-liquid flows in horizontal ducts. In this case, an initially flat granular bed may be unstable, generating ripples and dunes. This three-phase flow, although complex, can be modeled under some simplifying assumptions. This paper presents a model for the estimation of some bed-load characteristics. Based on parameters easily measurable in industry, the model can predict the local bed-load flow rates and the celerity and the wavelength of instabilities appearing on the granular bed.
\end{abstract}

\begin{keyword}
%% keywords here, in the form: keyword \sep keyword
Closed-conduit flow \sep pressure gradient \sep sediment transport \sep bed-load \sep instability
%% MSC codes here, in the form: \MSC code \sep code
%% or \MSC[2008] code \sep code (2000 is the default)

\end{keyword}

\end{frontmatter}

%%
%% Start line numbering here if you want
%%
% \linenumbers

%% main text
%\section{}
%\label{}

\section{INTRODUCTION}

Closed-conduit pressure-driven flows can entrain solid particles as a mobile granular bed, known as bed-load. Bed-load occurs when the shear stresses exerted by the fluid flow can displace some grains by rolling or sliding, but not as a suspension, forming then a mobile granular bed \cite{Bagnold_1, Bagnold_3, Raudkivi_1, Yalin_1}. A case frequently found in industry is the bed-load transport by stratified gas-liquid flows in horizontal ducts. Some examples are the stratified patterns appearing in the pumping of gas, oil and sand in petroleum pipelines and of gas, slurry and solid residues in sewage ducts.

In the presence of bed-load, the granular bed may be unstable, generating ripples and dunes. In a closed-conduit flow, these forms create supplementary pressure losses and pressure and flow rate transients \cite{Kuru, Franklin_3}, so that a better knowledge of this kind of transport is of great importance to improve related industrial processes.

\begin{figure}
  \begin{center}
    \includegraphics[width=0.85\columnwidth]{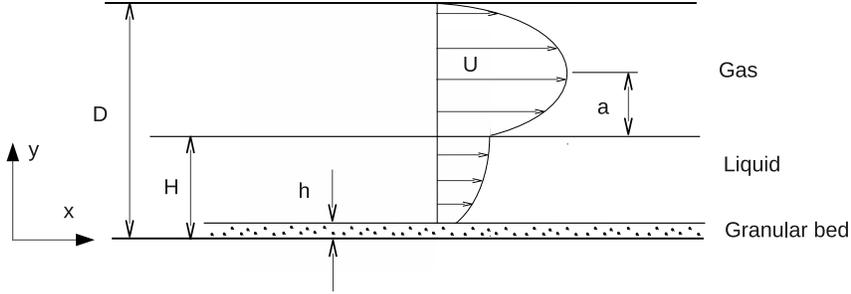}
    \caption{Granular bed sheared by a gas-liquid stratified flow. The thickness of the granular bed is $h$, that of the liquid layer is $H-h$, $D$ is the duct height, $U$ is the velocity of the mean flow in the longitudinal direction and $a$ is the distance from the maximum of the velocity profile to the gas-liquid interface.}
    \label{fig:escopo}
  \end{center}
\end{figure}

This three-phase flow, although complex, can be modeled under some simplifying assumptions. A model for the bed-load transport by stratified gas-liquid flows is proposed here for a case commonly found in industry, depicted in Fig. \ref{fig:escopo}. The flow is pressure-driven in horizontal ducts, where the mean thickness of the granular layer $h$ is many times smaller than that of the liquid layer $H-h$, and the mean thickness of the liquid layer $H-h$ is of the same order of magnitude as that of the gas $D-H$, i.e., $h/H\ll H/D=O(1)$. The thickness of the liquid layer is assumed to be much larger than the capillary length.

This paper presents a model valid for the specified case. The main purpose of the model is to predict, based on quantities easily measurable in industry, the local bed-load flow rates and the growth rate, the celerity and the wavelength of instabilities appearing on the granular bed.

The next section describes the physics and the main equations of the model. The following section describes the initial instabilities appearing on the granular bed and discusses their evolution to a saturated state. The conclusion section follows.

\section{BED-LOAD IN A STRATIFIED GAS-LIQUID FLOW}
\label{section:bedload}

The analyzed problem is very complex, so that the model's main objective is the estimation of the general behavior of the granular bed, adding new insights to it. In this case, the consideration of a three-dimensional geometry does not assure a better understanding of the problem: in a three-dimensional model, the obtained scaling laws and equations are more complex and depend on particular geometries, lacking generality. For this reason, the proposed model is two-dimensional.

The model divides the flow into a basic state, where the flow properties are homogeneous and steady in time, and a perturbation, which includes all the deviations from the basic state. This is described next.

\subsection{Basic State}

The flow can be divided in three layers, corresponding to the gas phase, to the liquid phase and to the granular bed. The basic state is considered as a steady state flow in which the thickness of each layer is constant in space and time. In the following, the variables in the basic state are identified by the subscript $0$. 

Equations relating interface shear stresses and pressure gradients can be found for the gas and liquid flows by integration of the momentum equation of the mean flow, as done by Cohen and Hanratty (1968) \cite{Cohen_Hanratty}, for example. Then, a semi-empirical equation can be employed to estimate the bed-load flow rate.

Considering the basic state as a fully-developed incompressible flow, the integration of the $y$ component of the mean flow momentum equation shows that the longitudinal pressure gradient does not vary in the $x$ direction, in both the gas and liquid layers. In the gas, the momentum equation of the mean flow in the $x$ direction

\begin{equation}
0\,=\,-\frac{1}{\rho}\frac{dP}{dx}\,-\,\frac{d\overline{u'v'}}{dy}\,+\,\nu \frac{d^2U}{dy^2}
\label{eq:qdm_x}
\end{equation}

\noindent can be integrated from $y=H_0$ to $y$, noting that at $y=H_0$ the interfacial stress is $\tau_{i,0}\,=\,\rho (-\overline{u'v'}\,+\,\nu dU/dy)$, and that at $y=(H_0+a_0)$ the velocity reaches its maximum value, so that $dU/dy=0$ and $\overline{u'v'}=0$. In Eq. \ref{eq:qdm_x}, $\rho$ and $\nu$ are, respectively, the specific mass and the kinematic viscosity of the fluid, $U$ is the velocity of the mean flow in the longitudinal direction, $u'$ and $v'$ are, respectively, the $x$ and the $y$ components of the velocity fluctuations, and $\overline{u'v'}$ is the stress tensor in the $xy$ plane. The integration yields

\begin{equation}
\tau_{i,0}\,=\,a_0\left(-\frac{dp}{dx}\right)
\label{eq:stress_gas}
\end{equation}

In the liquid, Eq. \ref{eq:qdm_x} is integrated from the granular bed $y=h_0$, where the shear stress $\tau_0$ is of viscous nature due to the the small values of fluctuations in this region, to the gas-liquid interface $y=H_0$. Considering that $O(h_0)<O(H_0)$, the integration yields

\begin{equation}
\tau_{0}\,=\,(a_0+H_0)\left(-\frac{dp}{dx}\right)
\label{eq:stress_liquid}
\end{equation}

Equation \ref{eq:stress_liquid} relates the shear stress on the granular bed to the pressure gradient $(dp/dx)$ of the flow and to the position of the maximum $y=a_0$ of the velocity profile. These quantities are easily measurable, or can be estimated.

The pressure gradient can be measured by pressure transducers installed along the flow-line. However, if they are absent, the pressure gradient may be estimated by the Lockhart-Martinelli correlations for gas-liquid flows \cite{Martinelli, Lockhart_Martinelli}. These correlations, based on both the flow rates of the fluids and the gas fraction, were initially proposed for separated horizontal flows without phase change or significant acceleration. For this reason, they are applied here to horizontal stratified flows. In addition, they are relatively simple and have been largely employed and tested \cite{Wallis}. Basically, this method allows the estimation of the pressure gradient $dp/dx$ for gas-liquid flows from the pressure gradient $(dp/dx)_g$ that would exist if pure gas flowed at the same mass flow rate

\begin{equation}
\frac{dp}{dx}\,=\,\phi_g^2\left(\frac{dp}{dx}\right)_g
\label{eq:lockhart_martinelli}
\end{equation}

\noindent where the multiplied factor $\phi_g^2$ may be determined from the mean gas fraction $\alpha$ in the duct. Chisholm (1967) \cite{Chisholm} proposed, for both phases in turbulent regime,

\begin{equation}
\phi_g^2\,=\,1\,+\,20\left(\alpha^{-2.695}-1\right)^{1.25}\,+\,\left(\alpha^{-2.695}-1\right)^{2.5}
\label{eq:phi_g}
\end{equation}

\noindent Equation \ref{eq:phi_g} is valid for $Re_{ph}>10^3$, where $Re_{ph}=\rho_{ph} j_{ph}D/\mu_{ph}$ is the Reynolds number of one phase, $j_{ph}=\dot{\forall}_{ph}/A_D$ is the volumetric flux, $\dot{\forall}_{ph}$ is the volumetric flow rate, $A_D$ is the duct cross-section, $\mu_{ph}$ is the dynamic viscosity and the subscript $ph$ indicates the phase.

The position of the maximum $y=a_0$ of the gas flow can be estimated as $a_0\approx 0.5(D-H_0)$. For example, Cohen and Hanratty (1968) \cite{Cohen_Hanratty} found $0.4(D-H_0)<a_0<0.6(D-H_0)$ in their experiments.

In the basic state, the bed-load flow rate and the fluid flow are in equilibrium. In this case, called \textit{saturated}, a semi empirical equation for the bed-load flow rate may be employed. Meyer-Peter and Mueller (1948) \cite{Mueller} performed exhaustive experiments in water flumes and obtained a widely used equation that is employed here

\begin{equation}
\frac{Q_{0}}{\sqrt{(S-1)gd^3}}\,=\,8\left(\theta_0-\theta_{th}\right)^{3/2}
\label{eq:mueller}
\end{equation}

\noindent where $Q_{0}$ is the bed-load flow rate, by unit of width, on the basic state, $S=\rho_s/\rho_l$ is the ratio of the specific masses of the solid $\rho_s$ and the liquid $\rho_l$, $d$ is the mean grain diameter, $g$ is the gravitational acceleration and $\theta_0$ is the Shields number in the basic state.

The Shields number $\theta$ is a dimensionless parameter characterizing the bed-load, defined as the ratio of the entraining force, scaling as $\tau_{0} d^{2}$, to the resisting force, that scales with $(\rho_{s} - \rho_l)gd^{3}$

\begin{equation}
\theta = \frac{\tau}{(\rho_{s}-\rho_l)gd}
\label{shields}
\end{equation}

\noindent Bed-load takes place for $0.01\,\lesssim\,\theta\,\lesssim\,1$. The term $\theta_{th}$ corresponds to the Shields parameter at the bed-load threshold \cite{Yalin_1, Buffington_1, Charru_1}, and its value can be obtained by graphical charts or correlations \cite{Whitehouse_Hardisty, Soulsby_Whitehouse}.

\subsection{Perturbation}

In the presence of bed-load, instabilities may appear on the surface of the granular bed, perturbing the fluid flow. The instabilities are bed undulations, initially of small aspect ratio, that grow and generate ripples and dunes \cite{Engelund_Fredsoe}. Equations \ref{eq:stress_liquid} and \ref{eq:mueller}, obtained for the basic state, must then be corrected to take into account the perturbations. In the following, variables with the subscript $pert$ are related to deviations from the basic state, and the ones without subscript are related to local values, i.e., the basic state added to the perturbation.

Franklin (2012) \cite{Franklin_6} obtained an expression for the shear stress on an undulated granular bed in the case of gravitational free-surface flows. His model supposed the existence of an outer region, far enough from the bed, where the perturbations induced by the bedforms had a relatively small length-scale, so that the turbulence could not adapt to the mean flow. With this assumption, the perturbation of the Reynolds stresses are negligible in this region and a potential solution shall exist at the leading order. Near the bed, on the other hand, the viscous effects and the shear stresses are important, however the flow perturbations are driven mainly by the pressure field of the outer region. 

In the present model, it is assumed that the flow is turbulent and that $h/H\ll H/D=O(1)$, therefore an outer region is expected to exist. Near the bed the perturbations are driven by the pressure gradient of the outer region, so that the solutions obtained by \cite{Franklin_6} are valid and may be adapted to pressure driven flows. This is done next.

The shear stress on the undulated bed can be written as

\begin{equation}
\tau\,=\,\tau_0\left( 1+\tau_{pert}\right)
\label{eq:total_stress}
\end{equation}

\noindent where $\tau_0$ is the shear stress on a flat bed (basic state) and $\tau_{pert}$ is the perturbation of the shear stress caused by the bed undulation. The shear stress perturbation $\tau_{pert}$ was found by \cite{Franklin_6} as

\begin{equation}
\tau_{pert}=f(Fr)A\left(\frac{h}{H_0}\,+\,B\partial_{x}h\right)
\label{eq:tau_real}
\end{equation}

\noindent where  $A=O(1)$ and $B=O(0.1)$ are constants and $f(Fr)$ is 

\begin{equation}
f(Fr)=\frac{-1}{Fr^2-1}
\label{eq:C_1}
\end{equation}

\noindent  where $Fr=U_0/\sqrt{gH_0}$ is the Froude number (the ratio between the velocity of the mean flow and the celerity of surface gravity waves) and $U_0$ is a characteristic velocity of the mean flow, considered here as the mean of the mean flow velocity profile, at the basic state. Equation \ref{eq:C_1} shows that $f(Fr)$ is constant for a given fluid flow.

With the shear stress given by Eq. \ref{eq:total_stress}, the saturated bed-load flow rate at the local flow conditions $q_{sat}$ (by unit of width) can be computed employing the Meyer-Peter and Mueller (1948) \cite{Mueller} equation

\begin{equation}
\frac{q_{sat}}{\sqrt{(S-1)gd^3}}\,=\,8\left(\theta-\theta_{th}\right)^{3/2}
\label{eq:mueller_local}
\end{equation}

\noindent which is similar to Eq. \ref{eq:mueller}, but for the local conditions. However, the shear stress caused by the fluid on an undulated bed varies in space, as showed by Eqs. \ref{eq:total_stress} and \ref{eq:tau_real}. Due to the inertia of the grains, the bed-load flow rate lags some distance (or time) to adapt to the local conditions of the fluid flow. This distance is a characteristic length called \textit{saturation length}, $L_{sat}$. It was showed by \cite{Charru_3, Franklin_4} that, in the case of liquid turbulent flows,

\begin{equation}
L_{sat}=\frac{u_*}{U_s}d
\label{eq:lsat}
\end{equation}

\noindent where $u_*=\sqrt{\tau_0/\rho_l}$ is the shear velocity and $U_s$ is the settling velocity of one grain. A simplified expression taking into account this relaxation effect can be obtained from the erosion-deposition model of \cite{Charru_1}

\begin{equation}
\frac{\partial q}{\partial x}\,=\,\frac{q_{sat}-q}{L_{sat}}
\label{eq:relax}
\end{equation}

\noindent where $q$ is the bed-load flow rate, by unit of width, on the undulated bed.

\subsection{Bed-load estimation on pressure-driven gas-liquid flows}

The local values of the bed-load flow rate can be computed employing the described equations. From the delivered gas and liquid flow rates, the pressure gradient can be estimated by Lockhard-Martinelli correlations (Eqs. \ref{eq:lockhart_martinelli} and \ref{eq:phi_g}). In cases where pressure transducers are installed along the line, the pressure gradient can be obtained directly from the measurements.  In the basic state, the shear stress on the granular bed and the bed-load flow rate can be computed with Eqs. \ref{eq:stress_liquid} and \ref{eq:mueller}, respectively. If the granular bed is perturbed, the shear stress on the undulated bed can be computed with Eqs. \ref{eq:total_stress} and \ref{eq:tau_real} and the bed-load flow rate can be estimated with Eqs. \ref{eq:mueller_local} and \ref{eq:relax}. In the latter case, however, computations require the form of the bed, i.e., the knowledge of the wavelength and of the amplitude of the bedforms. The determination of these values is presented in section \ref{section:stability}. Next, assuming that the form of the bed is known, an example of calculation of the local bed-load flow rate is made.

\begin{figure}
   \begin{minipage}[c]{.49\linewidth}
    \begin{center}
      \includegraphics[scale=0.45]{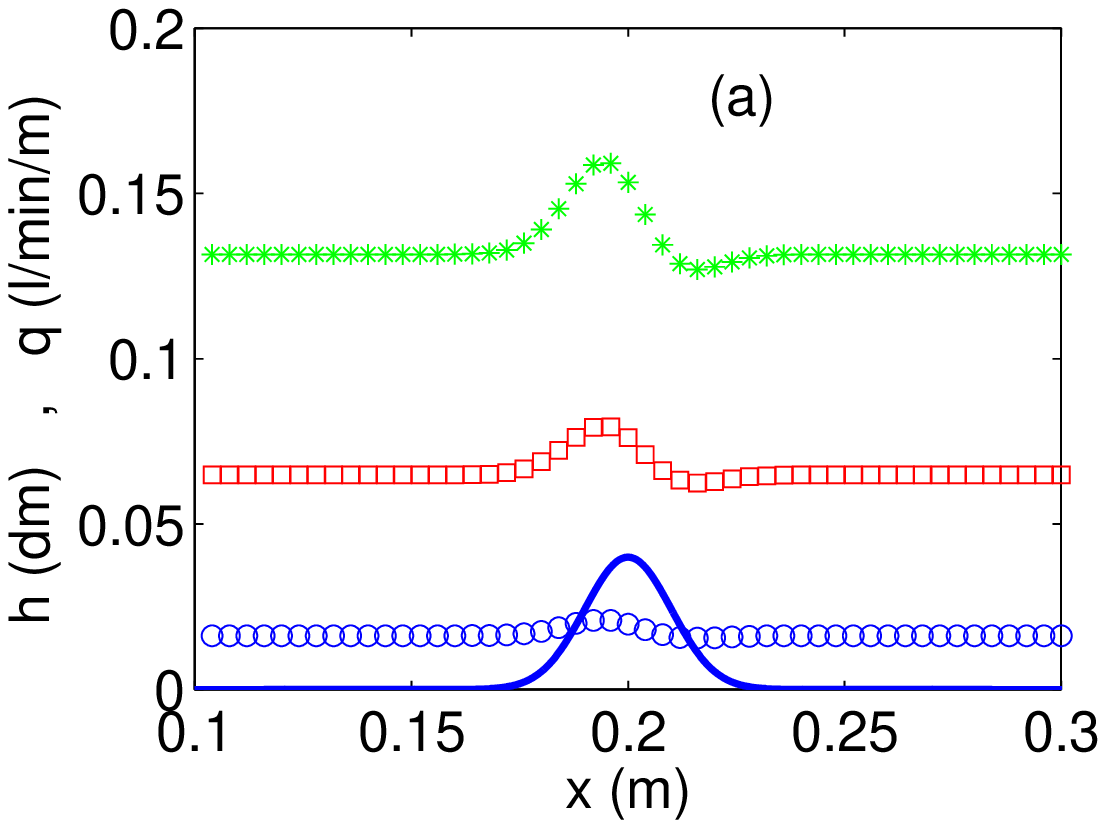}
    \end{center}
   \end{minipage} \hfill
   \begin{minipage}[c]{.49\linewidth}
    \begin{center}
      \includegraphics[scale=0.45]{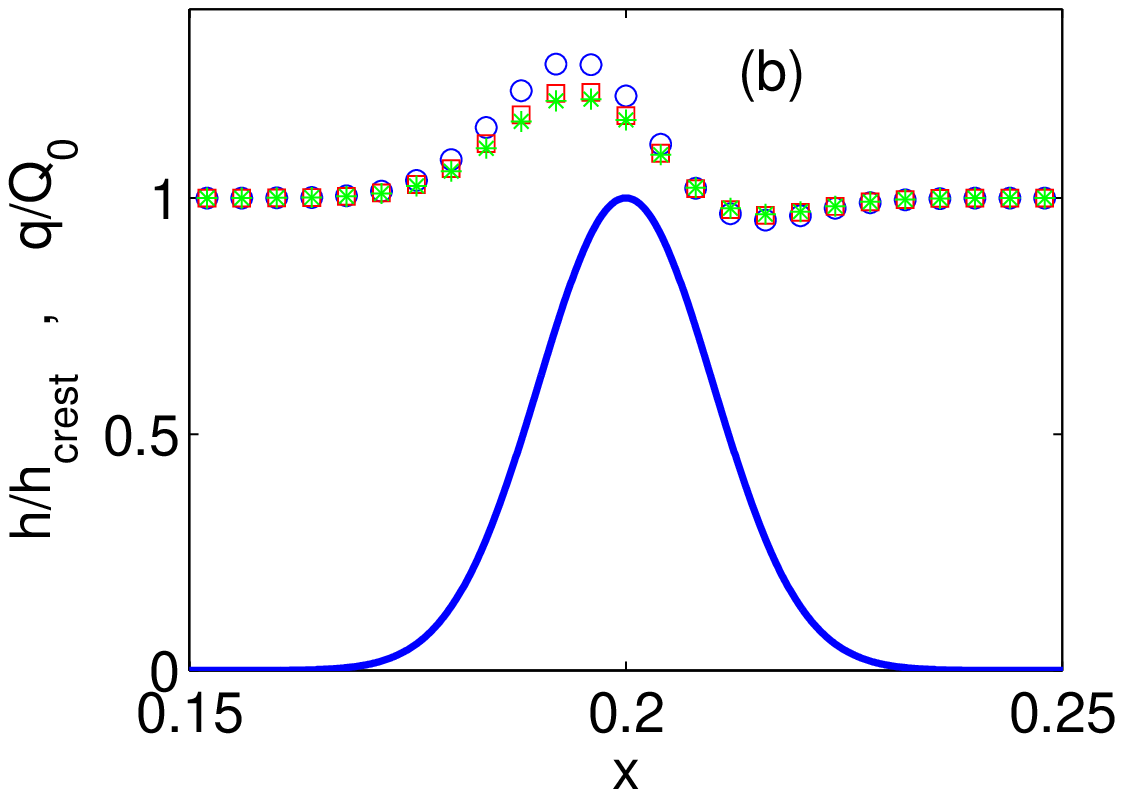}
    \end{center}
   \end{minipage}
\caption{(a) Bedform height and bed-load flow rate as a function of the longitudinal position $x$. (a) Dimensional form and (b) dimensionless form. The continuous line correspond to the bedform height and the symbols to the bed-load flow rate. The circles correspond to $dp/dx=-4.3Pa/m$. The squares and asterisks correspond to $2dp/dx$ and $3dp/dx$, respectively.}
\label{fig:bedload}
\end{figure}

Bed-load flow rates were estimated employing the following parameters: $d=0.25mm$, $\rho_l=10^3kg/m^3$, $\mu_l=10^{-3}Pa.s$, $\rho_g=1.2kg/m^3$, $\mu_g=2.10^{-5}Pa.s$, $\rho_s=2600kg/m^3$, $D=0.15m$, $H_0=0.35D$, $a_0=0.55(D-H_0)$ and $U_s=0.01m/s$. The total longitudinal domain was $0.4m$ and the bedform was approximated by a Gaussian function, with an aspect ratio of $0.1$, mean at $x=0.2m$ and standard deviation of $S_{d}=0.01m$. The value of the length of the bedform is then $L\approx 4S_{d}=0.04m$, which is determined by the stability analysis of section \ref{section:stability} (cf. Tab. \ref{tab:inst}). Figure \ref{fig:bedload} summarizes the results of these computations, presenting the bed-load flow rate as a function of the longitudinal position $x$. 

Figure \ref{fig:bedload}(a) shows the bed-load flow rate $q$ and the bedform height $h$ as a function of the longitudinal position $x$, for three different pressure gradients. The continuous line corresponds to the bedform height $h$ and the symbols to the bed-load flow rate. The circles correspond to $dp/dx=-4.3Pa/m$, estimated from $j_l=0.1m/s$, $j_g=3m/s$ and $\alpha=0.65$. The squares and asterisks correspond to $2dp/dx$ and $3dp/dx$, respectively.

Far from the bedform, the bed-load flow rate is near the basic state and it is expected to vary as $(dp/dx)^{(3/2)}$. This is shown in the regions far from the crest of Fig. \ref{fig:bedload}(a), where the values of $q$ do not vary with $x$. In these regions ($x<0.15m$ or $x>0.25m$) we find effectively that $q\,\sim\,(dp/dx)^{(3/2)}$. Over the bedforms, the variation of $q$ depends also on the form of the bed because the fluid flow, which entrains the grains, is perturbed by it. This perturbation is taken into account via Eqs. \ref{eq:total_stress} and \ref{eq:tau_real}, so that the relation between the bed-load and the pressure gradient in this region depends on the local slope of the bed.

Figure \ref{fig:bedload}(b) presents the bed-load flow rate normalized by its value at the basic state $q/Q_0$ as a function of the longitudinal position $x$. The bedform height normalized by its maximum $h/h_{crest}$ is also shown (continuous line). The symbols are the same as that of Fig. \ref{fig:bedload}(a). From Fig. \ref{fig:bedload}(b), it is clear that far from the bedform crest, where the bed is nearly flat, the bed-load flow rate is close to its saturated value $Q_0$.

Figure \ref{fig:bedload}(b) also shows that the maximum of the bed-load flow rate occurs upstream of the bedform crest, which characterizes an unstable situation \cite{Franklin_4}: deposition occurs at the crest and the bedform amplitude tends to increase. This agrees with the analysis described in section \ref{section:stability}, as the bedform length employed in computations was chosen as the most unstable mode (cf. Tab. \ref{tab:inst}). Just downstream of the bedform crest, due to smaller shear stresses in this region, the bed-load flow rate is smaller than the saturated value. This indicates that the perturbation caused by the imposed Gaussian form tends to accumulate grains at the crest and just downstream of it, so that the bedform tends to assume a triangular form with a gentle upstream slope and a rather abrupt downstream slope, just as observed in nature \cite{Bagnold_1, Yalin_1, Raudkivi_1, Raudkivi_2}.

Table \ref{tab:bedload} shows, for the computations presented in Fig. \ref{fig:bedload}, the values of the bed-load flow rate at the basic state $Q_0$, the maximum local value of the bed-load flow rate $q_{max}$, the position where this maximum value occurs $x_{max}$, the value of the bed-load flow rate at the bedform crest $q_{crest}$, the minimum local value of the bed-load flow rate $q_{min}$ and the position where this minimum value occurs $x_{min}$, in dimensional form. We verify that the maximum of the bed-load flow rate occurs slightly upstream of the bedform crest, and that a minimum value is reached downstream of the crest.

The values presented in Tab. \ref{tab:bedload} can be employed as initial estimations in the design of gas-liquid flow lines conveying grains as bed-load. To the author's knowledge, this is the first time that a model is proposed to give bed-load estimations as a function of the pressure gradient, which is an easily measurable parameter.

\begin{table}[htbp]
\begin{center}
\begin{tabular}{|c|c|c|c|c|c|c|}
	\hline
	$-dp/dx$ & $Q_0$ & $q_{max}$ & $x_{max}$ & $q_{crest}$ & $q_{min}$ & $x_{min}$\\
	{\scriptsize $(Pa/m)$} & {\scriptsize $(l/min/m)$} & {\scriptsize $(l/min/m)$} & {\scriptsize $(m)$} & {\scriptsize $(l/min/m)$} & {\scriptsize $(l/min/m)$} & {\scriptsize $(m)$}\\
	\hline
	4.3 & $0.0163$ & $0.0210$ & $0.1940$ & $0.0198$ & $0.0155$ & $0.2158$\\
	\hline
	8.7 & $0.0649$ & $0.0797$ & $0.1942$ & $0.0762$ & $0.0624$ & $0.2160$\\
	\hline
	13.1 & $0.1316$ & $0.1597$ & $0.1944$ & $0.1533$ & $0.1270$ & $0.2163$\\
	\hline
\end{tabular}
\caption{Bed-load flow rate at the basic state $Q_0$, the maximum local value of the bed-load flow rate $q_{max}$, the position where this maximum value occurs $x_{max}$, the value of the bed-load flow rate at the bedform crest $q_{crest}$, the minimum local value of the bed-load flow rate $q_{min}$ and the position where this minimum value occurs $x_{min}$, for different pressure gradients $dp/dx$.}
\label{tab:bedload}
\end{center}
\end{table}

\section{STABILITY ANALYSIS}
\label{section:stability}

Fouri\`ere et al.(2010) \cite{Fourriere_1} proposed, in the case of river streams, that ripples are primary linear instabilities while dunes are formed from the coalescence of ripples. Given the scope of the present model, the fundamental idea of \cite{Fourriere_1} is employed here: ripples are bedforms whose wavelength does not scale with the flow depth and are formed as a primary linear stability, while dunes have a wavelength which scales with the flow depth and are formed as a secondary instability. The main difference here is that, given the scales of the liquid depth and that of ripples, dunes will be formed from a relative small quantity of ripples.

With this assumption, the initial instabilities give the scales of ripples. This is presented in subsection \ref{subsection:linear}, where a linear stability analysis is made based on the equations presented in section \ref{section:bedload}. Subsection \ref{subsection:sat} discusses the nonlinearities and the formation of dunes.

\subsection{Linear analysis}
\label{subsection:linear}

Franklin (2010) \cite{Franklin_4} presented a linear stability analysis of a granular bed sheared by a turbulent liquid flow, without free-surface effects. An analysis of the same kind is presented here, however, different from \cite{Franklin_4}, the effects of the free surface as well as the bed-load threshold are taken into account. Also, as the interest here is in pressure-driven flows, the obtained equations are analyzed in terms of pressure gradients.

The linear analysis is based on four equations. The first one is the perturbation of the fluid flow by the shape of the bed, given by Eq. \ref{eq:tau_real}, which in the Fourier space is

\begin{equation}
\tilde{\tau}_{pert}\,\approx\,f(Fr)A_2\tilde{h} \left[ |k|+iBk \right]
\label{eq:tau_fourier}
\end{equation}

\noindent as showed in \cite{Franklin_6}, where $k$ is the wavenumber in the longitudinal direction and the tilde denotes variables in the Fourier space.

The second equation is the saturated bed-load flow rate at the local flow conditions $q_{sat}$, given by Eq. \ref{eq:mueller_local}. Here, different from \cite{Franklin_4}, the flow is not assumed to be far from the threshold conditions ($\theta>>\theta_{th}$), so that the threshold term is conserved in the equation. Equation \ref{eq:mueller_local} can be linearized and made dimensionless by dividing it by a reference value, taken as the flow rate at the basic state. Considering the shear stress perturbation given by Eq. \ref{eq:tau_fourier}, 

\begin{equation}
\frac{q_{sat}}{Q_0}\,\sim\,D_1\,+\,\frac{3}{2}D_2f(Fr)A_2\left[ |k|+iBk \right]h
\label{eq:mueller_linearized}
\end{equation}

\noindent where

\begin{equation}
D_1\,=\,\left( 1-\frac{\tau_{th}}{\tau_0} \right)^{3/2}
\end{equation}

\begin{equation}
D_2\,=\,\left( 1-\frac{\tau_{th}}{\tau_0} \right)^{1/2}
\label{eq:d2}
\end{equation}

\noindent and $\tau_{th}$ is the shear stress corresponding to $\theta_{th}$. $D_1$ and $D_2$ are constants for a granular bed of a given granulometry under a given fluid flow. The great advantage of making the saturated bed-load flow rate dimensionless is to have an indication of how the normal modes shall be, as seen next.  

The third employed equation is Eq. \ref{eq:relax}, that accounts for the relaxation effects related to the transport of grains, and the fourth one is the mass conservation of granular matter

\begin{equation}
\frac{\partial h}{\partial t}\,+\,\frac{1}{\varphi}\frac{\partial q}{\partial x}\,=\,0
\label{eq:exner}
\end{equation}

\noindent where $t$ is the time and $\varphi$ is the solids concentration of the granular bed. Equations \ref{eq:relax} and \ref{eq:exner} are in the same form as in \cite{Franklin_4}.

Taking into account that the initial instabilities are small scale perturbations, solutions to Eqs. \ref{eq:relax}, \ref{eq:tau_fourier}, \ref{eq:mueller_linearized} and \ref{eq:exner} shall consider the bedform height $h$ and the bed-load flow rate $q$ as plane waves. Equation \ref{eq:mueller_linearized} indicates that $h$ and $q$ can be decomposed in normal modes of the form

\begin{equation}
h(x,t)\,=\,\xi e^{\sigma t-i\omega t+ikx}
\label{eq:normal_h}
\end{equation}

\begin{equation}
\frac{q(x,t)}{Q_0}\,=\,D_1+\gamma e^{\sigma t-i\omega t+ikx}
\label{eq:normal_q}
\end{equation}

\noindent where $\sigma$ is the growth rate and $\omega$ is the frequency. The insertion of the normal modes given by Eqs. \ref{eq:normal_h} and \ref{eq:normal_q} into Eqs. \ref{eq:relax}, \ref{eq:tau_fourier}, \ref{eq:mueller_linearized} and \ref{eq:exner} gives origin to an eigenvalue problem

\begin{equation}
\left[\begin{array}{cc}\sigma-i\omega & (1/\varphi)ikQ_0 \\ (3/2)D_2\left( A_2 |k|+iBk \right) & -(1+ikL_{sat}) \\ \end{array}\right]\left[\begin{array}{c}\xi \\ \gamma \\ \end{array}\right] \,=\,\left[\begin{array}{c}0 \\ 0 \\ \end{array}\right]
\label{eq:eigenvalue}
\end{equation}

\noindent where $A_2=f(Fr)A$ is constant for a given fluid flow. The non-trivial solution of Eq. \ref{eq:eigenvalue} gives the growth rate and the frequency of initial instabilities. The group velocity corresponds to $c_G\,=\,d\omega /dk$,

\begin{equation}
\sigma\,=\,\frac{3}{2}\frac{D_2Q_0}{\varphi}\frac{k^2\left( B-A_2|k|L_{sat}\right)}{1+\left( kL_{sat}\right)^2}
\label{eq:sigma}
\end{equation}

\begin{equation}
c_G\,=\,\frac{3}{2}\frac{D_2Q_0}{\varphi}\frac{2A_2|k|+BL_{sat}k^2\left( 3+\left(L_{sat}k\right)^2\right)}{\left( 1+\left( kL_{sat}\right)^2\right)^2}
\label{eq:cg}
\end{equation}

\begin{figure}
   \begin{minipage}[c]{.49\linewidth}
    \begin{center}
      \includegraphics[scale=0.45]{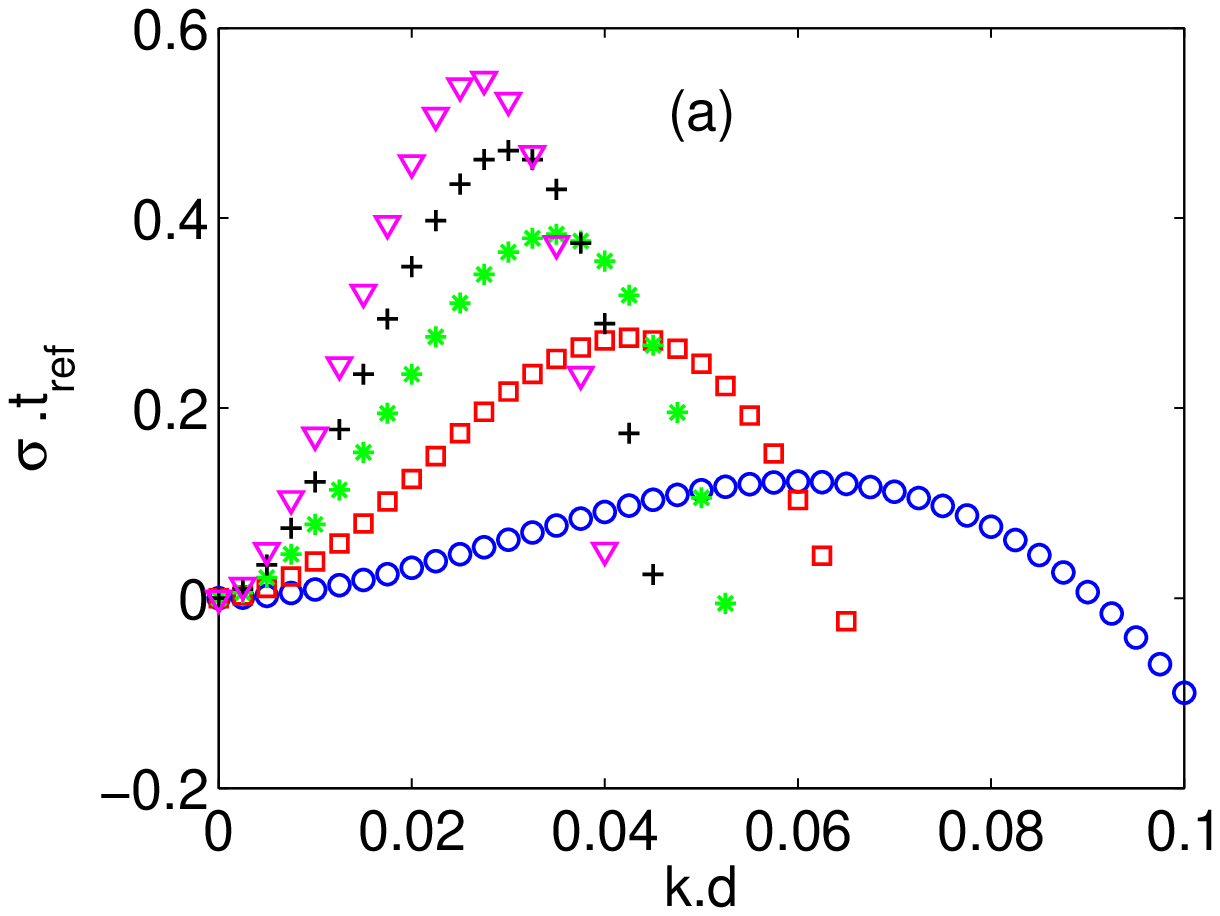}
    \end{center}
   \end{minipage} \hfill
   \begin{minipage}[c]{.49\linewidth}
    \begin{center}
      \includegraphics[scale=0.45]{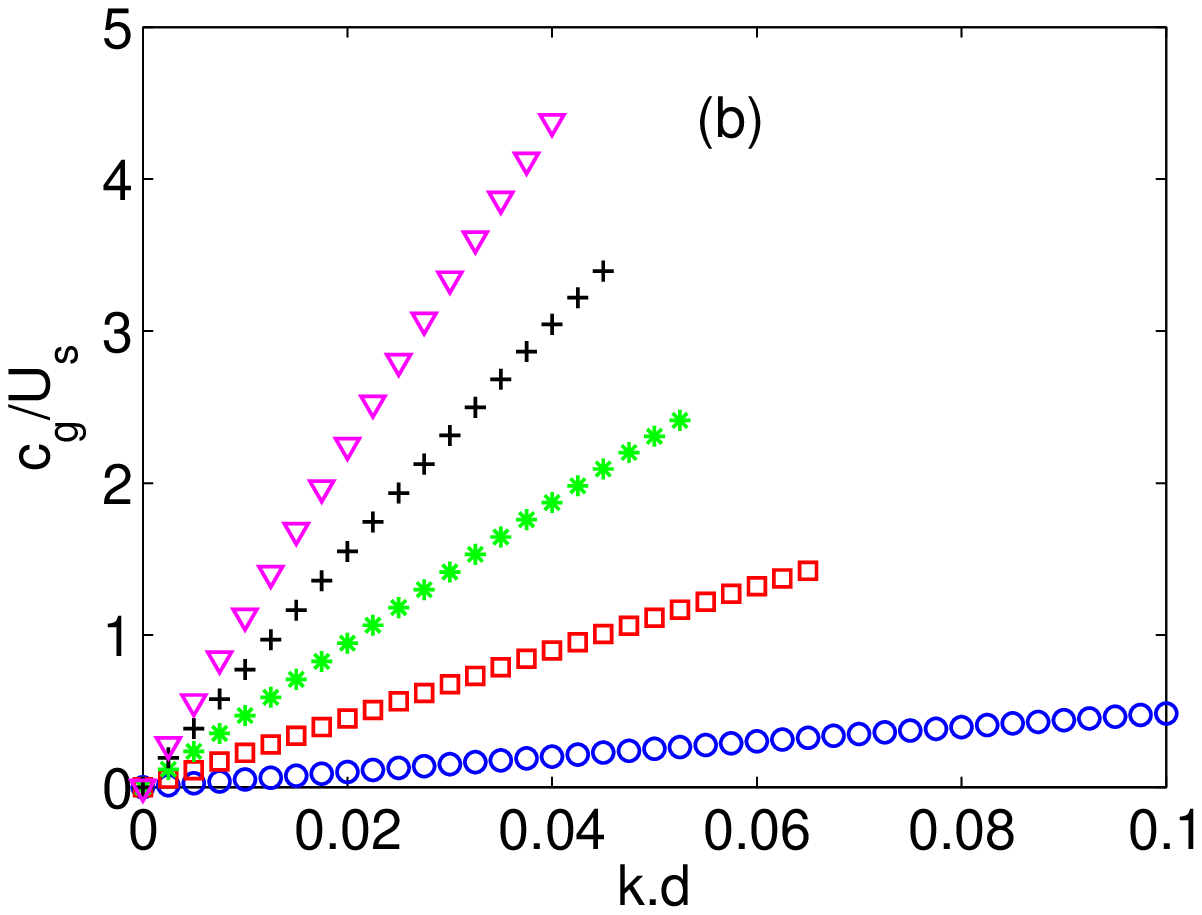}
    \end{center}
   \end{minipage}
\caption{(a) Dimensionless growth rate $\sigma t_{ref}$ and (b) dimensionless celerity $c_G/U_s$ of initial bedforms as functions of the dimensionless wave-number $kd$. The list of symbols is presented in Tab. \ref{tab:inst}.}
\label{fig:inst}
\end{figure}

The form of Eq. \ref{eq:sigma}, that can be seen in Fig. \ref{fig:inst}(a), shows that $\sigma$ has a maximum that corresponds to a most unstable mode, which can then be found from $d\sigma /dk=0$. In obtaining the most unstable mode, two approximations can be made: (i) the nature of the instability allows a long-wavelength approximation (Fig. \ref{fig:inst}), so that higher order terms in $k$ can be neglected; and (ii) the value of $A_2$ can be considered as constant because the initial instabilities always happens at $Fr=O(0.1)$ in the analyzed case. With these assumptions, the most unstable wavenumber $k_{max}$ is

\begin{equation}
k_{max}\,\approx\,\frac{2}{3}\frac{B}{A_2}\frac{1}{L_{sat}}
\label{k_max}
\end{equation}

\noindent so that the wavelength $\lambda_{max}$, the growth rate $\sigma_{max}$ and the celerity $c_{G,max}$ of the most unstable mode are

\begin{equation}
\lambda_{max}\,\approx\,\frac{3\pi A_2}{B}L_{sat}
\label{L_max}
\end{equation}

\begin{equation}
\sigma_{max}\,\approx\,\frac{2}{3}\frac{D_2Q_0}{\varphi}\frac{B^2}{A_2^2}\frac{1}{(L_{sat})^2}
\label{sigma_max}
\end{equation}

\begin{equation}
c_{G,max}\,\approx\,2B\frac{D_2Q_0}{\varphi}\frac{1}{L_{sat}}
\label{cg_max}
\end{equation}

As the growth rate is exponential in the liner phase of the instability (Eqs. \ref{eq:normal_h} and \ref{eq:normal_q}), the most unstable mode grows much faster than the others, so that it prevails. Initial bedforms appearing on the bed follow this mode and, as showed in \cite{Franklin_6, Franklin_5}, they saturate, keeping the same wavelength. In pressure driven flows, it is then interesting to know how these forms vary with the pressure gradient $dp/dx$. From Eqs. \ref{eq:mueller}, \ref{eq:lsat}, \ref{eq:d2}, \ref{L_max}, \ref{sigma_max} and \ref{cg_max}

\begin{equation}
\lambda_{max}\,\sim\,\left( -\frac{dp}{dx}\right)^{\left(1/2\right)}
\label{scaling:L_max}
\end{equation}

\begin{equation}
\sigma_{max}\,\sim\,\left( -\frac{dp}{dx}\right)^{\left(1/2\right)}
\label{scaling:sigma_max}
\end{equation}

\begin{equation}
c_{G,max}\,\sim\,\left( -\frac{dp}{dx}\right)
\label{scaling:cg_max}
\end{equation}

Figure \ref{fig:inst} presents the normalized growth rate $\sigma t_{ref}$ and the normalized celerity $c_G/U_s$ of the initial bedforms as functions of the normalized wave-number $kd$, where $t_{ref}=d/U_s$ is the reference settling time. These values were computed from Eqs. \ref{eq:sigma} and \ref{eq:cg} for five different pressure gradients: the circles correspond to $dp/dx=-4.3Pa/m$, estimated from $j_l=0.1m/s$, $j_g=3m/s$ and $\alpha=0.65$. The squares, asterisks, crosses and triangles correspond to $2dp/dx$, $3dp/dx$, $4dp/dx$ and $5dp/dx$, respectively. The values of all other parameters, except the bedform, whose scales shall be found from this stability analysis, are the same as that employed in section \ref{section:bedload}.

Figure \ref{fig:inst}(a) shows that the large wave-numbers are stable and the small ones unstable, corresponding then to a long-wave instability. In the unstable region, the growth rate presents a maximum at a wave-number characterizing then the most unstable mode. The maxima of $\sigma$ and the corresponding $k$ and $c_G$ were found from figure \ref{fig:inst}, and they were fitted as functions of $dp/dx$. The values found for the exponents of $-dp/dx$ related to $\lambda$, $\sigma$ and $c_G$ were $0.5$, $0.6$ and $1.1$, respectively, showing a good agreement with Eqs. \ref{scaling:L_max} to \ref{scaling:cg_max}. This corroborates the long-wavelength assumption made in finding these equations.

Table \ref{tab:inst} presents, for the most unstable mode, the wavelength $\lambda_{max}$, the celerity $c_{G,max}$ and the growth rate $\sigma_{max}$ in dimensional form, for each pressure gradient $dp/dx$. These are the values expected to prevail in the linear phase of the instability. As we will see in the next subsection, this wavelength persists in the nonlinear phase, so that it can be employed in the estimation of the local bed-load flow rate by the method presented in section \ref{section:bedload}.

\begin{table}[htbp]
\begin{center}
\begin{tabular}{|c|c|c|c|c|}
	\hline
	$-dp/dx$ & Symbol & $\lambda_{max}$ & $c_{G,max}$ & $\sigma_{max}$\\
	{\scriptsize $(Pa/m)$} & {\scriptsize $\cdots$} & {\scriptsize $(m)$} & {\scriptsize $(m/s)$} & {\scriptsize $(1/s)$}\\
	\hline
	4.3 & $\circ$ & $0.026$ & $0.003$ & $0.02$\\
	\hline
	8.7 & $\square$ & $0.037$ & $0.010$ & $0.05$\\
	\hline
	13.1 & $*$ & $0.045$ & $0.017$ & $0.07$\\
	\hline
	17.4 & $+$ & $0.052$ & $0.023$ & $0.09$\\
	\hline
	21.8 & $\nabla$ & $0.057$ & $0.031$ & $0.10$\\
	\hline
\end{tabular}
\caption{Wavelength $\lambda_{max}$, celerity $c_{G,max}$ and growth rate $\sigma_{max}$ of the most unstable mode for each pressure gradient $dp/dx$. The symbols employed in Fig. \ref{fig:inst} are also listed.}
\label{tab:inst}
\end{center}
\end{table}

To the author's knowledge, this is the first time that a model allows estimations of instabilities parameters as functions of the pressure gradient, which is an easily measurable quantity.

There is a lack of experiments in the scope of this model, even if there is a large number of industrial applications. However, the present model can be compared with experimental data of pressure driven liquid flows carrying grains as bed-load. Although the cases are different, the initial instabilities may scale in a similar manner, given that the initial bedforms have small amplitudes and are not expected to be affected by the presence of a free surface. There are at least two experimental works on bed instabilities under pressure driven closed-conduit flows: Kuru et al. (1995) \cite{Kuru} and Franklin (2008) \cite{Franklin_3}. Kuru et al. (1995) \cite{Kuru} performed experiments on a $7m$ long, $31.1mm$ diameter horizontal pipe, and employed mixtures of water and glycerin as the fluid media and glass beads as the granular media. Franklin (2008) \cite{Franklin_3} performed experiments on a $6m$ long, horizontal closed-conduit of rectangular cross-section ($120mm$ wide by $60mm$ high), made of transparent material, and employed water as the fluid and glass and zirconium beads as the granular media. In both works, the authors measured the wavelengths of the initial bedforms appearing on the granular bed. Given the small time scales of the problem and the presence of high uncertainties, the celerity and the growth rate were not reported.

\begin{figure}
  \begin{center}
    \includegraphics[width=0.50\columnwidth]{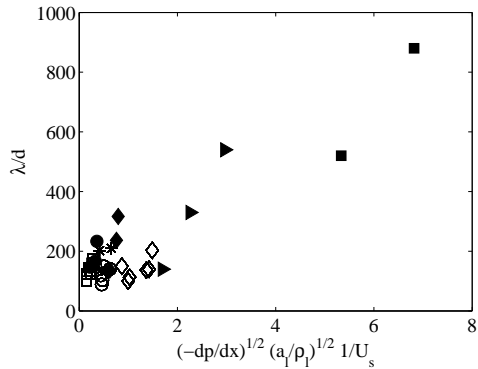}
    \caption{Dimensionless wavelength $\lambda/d$ as a function of the dimensionless square root of the pressure gradient $(-dp/dx)^{1/2}(a_l/\rho_l)^{1/2}(1/U_s)$. Filled circles, lozenges, triangles and squares correspond to $d=0.3mm $ and $\mu_l=1cP$, $d=0.3mm$ and $\mu_l =2.2cP$, $d=0.1mm$ and $\mu_l=1cP$ and $d=0.1mm$ and $\mu_l=2.1cP$, respectively (experimental data of Kuru et al., 1995 \cite{Kuru}). Open lozenges, circles, squares and asterisks correspond to $d=0.12mm$, $d=0.20mm$ and $d=0.50mm$ glass beads (in water) and to $d=0.19mm$ zirconium beads (in water), respectively (experimental data of Franklin, 2008 \cite{Franklin_3})}
    \label{fig:exper_data}
  \end{center}
\end{figure} 

The results of both works are summarized in Fig. \ref{fig:exper_data}. In order to directly compare the experimental data with the present model, Fig. \ref{fig:exper_data} presents the dimensionless wavelength $\lambda/d$ of initial ripples as a function of the dimensionless square root of the pressure gradient $(-dp/dx)^{1/2}(a_l/\rho_l)^{1/2}(1/U_s)$, where $a_l$ is the distance from the maximum of the liquid velocity profile to the bed. Filled symbols correspond to the experimental data of Kuru et al. (1995) \cite{Kuru} and open symbols to the experimental data of Franklin (2008) \cite{Franklin_3}. The description of each symbol is in the legend of Fig. \ref{fig:exper_data}.

The dimensionless parameters of Fig. \ref{fig:exper_data} come from Eqs. \ref{eq:stress_liquid}, \ref{eq:lsat} and \ref{L_max}, considering that $u_*=\sqrt{\tau_0/\rho_l}$ and changing $(a_0+H_0)$ by $a_l$. In this case, we find that

\begin{equation}
\frac{\lambda}{d}\,\sim\,\left(-\frac{dp}{dx}\right)^{1/2}\left(\frac{a_l}{\rho_l}\right) ^{1/2}\frac{1}{U_s}
\end{equation}

If we take into account the relatively high uncertainties, often present in measurements of bed instabilities, the alignment of the experimental data in Fig. \ref{fig:exper_data} seems to support the results of the proposed model, that the wavelength of initial bedforms varies as $(-dp/dx)^{1/2}$, even if the experimental data were obtained for a case different from the scope of the model. This gives some confidence in the proposed model.

\subsection{Considerations about nonlinearities}
\label{subsection:sat}

After the initial linear growth, bedforms attenuate their growth rate while keeping the same wavelength \cite{Franklin_5}. In the case of river streams, Franklin (2011) \cite{Franklin_6} showed that the initial bedforms saturate and generate ripples, which then coalesce and generate larger forms. These larger forms grow until the free surface is locally perturbed and the subcritical-supercritical transition is reached, so that the water stream becomes a stable mechanism. These forms, called dunes, maintain a wavelength in the range $H_0\,\lesssim\,\lambda\,\lesssim\,10H_0$ and are the result of a secondary instability, while ripples result from a primary instability.

The same reasoning can be applied to the present case. Given the dimensions of the problem it is expected that ripples saturate and that dunes are formed from their coalescence. The main difference in the present problem is that the number of coalesced ripples forming a dune is smaller than in rivers, and, depending on the flow depth, dunes can even be formed from a single ripple. In all cases, the dune length scales as

\begin{equation}
\lambda_{dune}\,\sim\,H_0
\end{equation}

Measured values indicate that $H_0\,\lesssim\,\lambda\,\lesssim\,10H_0$ \cite{Guy, Allen, Julien, Coleman_3} in river flows. The celerity of dunes can be obtained from the mass conservation of grains (Eq. \ref{eq:exner}) together with Eq. \ref{eq:mueller_local}. This gives an advection equation equal to the one obtained in \cite{Franklin_6}, however with a different value for the celerity

\begin{equation}
c_{dune}\,\approx\,-\frac{24}{H_0(S-1)g}\left(\tau_0\right)^{\frac{3}{2}}\frac{1}{Fr^2-1}
\label{eq:cdune}
\end{equation}

\noindent and then, from Eq. \ref{eq:stress_liquid}, the celerity of dunes scales as

\begin{equation}
c_{dune}\,\sim\,\left( -\frac{dp}{dx}\right)^{(3/2)}
\label{eq:var_cdune}
\end{equation}

The model then predicts the coexistence of two different types of bedforms. The smaller type corresponds to ripples, whose length scales as $\lambda_{ripple}\,\sim\,\left( -dp/dx\right)^{\left(1/2\right)}$ and whose celerity scales as $c_{ripple}\,\sim\,\left( -dp/dx\right)$. The other type corresponds to dunes, that are larger forms scaling as $\lambda_{dune}\,\sim\,H_0$ and $c_{dune}\,\sim\,\left( -dp/dx\right)^{(3/2)}$. Depending on the flow depth, the length of saturated bedforms will obey Eq. \ref{L_max} in subcritical flow, or $H_0\,\lesssim\,\lambda\,\lesssim\,10H_0$ otherwise. In any case, the bedforms grow until their crests reach a height corresponding to a local Froude number near the transition $Fr\approx 1$. This picture predicts that at development regions, such as the duct entrance, the granular bed and the fluid flow are adapting themselves so that ripples will be formed and predominate. In regions of fully-developed flow, the time and length scales are large enough to allow the growth of ripples and their coalescence, so that dunes will predominate.

Once determined the length and the amplitude of the bedforms, the local bed-load flow rates may be estimated by the method described in section \ref{section:bedload}. This justifies \textit{a posteriori} the scales employed in the estimations of bed-load flow rates presented in section \ref{section:bedload}.

\section{CONCLUSIONS}
\label{section:conclusions}

This paper presented a model for the estimation of bed-load and associated instabilities in stratified gas-liquid flows. The model focuses on the case of turbulent pressure-driven stratified flows, in horizontal closed-conduits. It divides the fluid flow and the bed-load into a flat basic state, and undulated perturbations. In order to compute the perturbations, the scales of the bed undulations are obtained by a stability analysis. The local shear stresses and bed-load flow rates are then the sum of the basic state and the perturbations.

The stability analysis predicted the coexistence of two different types of bedforms: the ripples, that are primary instabilities formed from the initial bedforms, and the dunes, that are secondary instabilities formed from the coalescence of ripples. Ripples predominate in development regions, such as the duct entrance, while dunes predominate in fully-developed regions of the flow.

The model can be employed to estimate bed-load flow rates and the shape of the bed in the design of lines of pressure-driven gas-liquid flows conveying grains. Other than the fluids and grains properties and the main geometry of the flow, the model needs the pressure gradient as an input. This quantity can be easily estimated or measured. To the author's knowledge, this is the first time that a model for a three-phase flow allows, from pressure gradient measurements, the estimation of instabilities parameters and bed-load flow rates.

\section{ACKNOWLEDGMENTS}

The author is grateful to Petrobras S.A. (contract number 0050.0045763.08.4) and to FAEPEX/UNICAMP (conv. 519.292, project 1435/12).

%% The Appendices part is started with the command \appendix;
%% appendix sections are then done as normal sections
%% \appendix

%% \section{}
%% \label{}

%% References
%%
%% Following citation commands can be used in the body text:
%% Usage of \cite is as follows:
%%   \cite{key}         ==>>  [#]
%%   \cite[chap. 2]{key} ==>> [#, chap. 2]
%%

%% References with bibTeX database:

%\bibliographystyle{elsarticle-num}
%\bibliography{<your-bib-database>}

\bibliography{references}
\bibliographystyle{elsart-num}

%% Authors are advised to submit their bibtex database files. They are
%% requested to list a bibtex style file in the manuscript if they do
%% not want to use elsarticle-num.bst.

%% References without bibTeX database:

% \begin{thebibliography}{00}

%% \bibitem must have the following form:
%%   \bibitem{key}...
%%

% \bibitem{}

% \end{thebibliography}

\end{document}